\documentclass[prc,twocolumn,showpacs]{revtex4}
\usepackage{graphicx}
\usepackage{dcolumn}
\usepackage{bm}

\begin{document}

\title{Shape evolution in Yttrium and Niobium neutron-rich isotopes}

\author{R. Rodriguez-Guzman$^{1}$, P. Sarriguren$^{1}$}
\author{L.M. Robledo$^{2}$}
\affiliation{
$^{1}$ Instituto de Estructura de la Materia, CSIC, Serrano
123, E-28006 Madrid, Spain \\
$^{2}$ Departamento  de F\'{\i}sica Te\'orica, M\'odulo 15,
Universidad Aut\'onoma de Madrid, 28049-Madrid, Spain
}

\date{\today}

\begin{abstract}

The isotopic evolution of the ground-state nuclear shapes and the
systematics of one-quasiproton configurations are studied in 
neutron-rich odd-$A$ Yttrium and Niobium isotopes. We use a
selfconsistent Hartree-Fock-Bogoliubov formalism based on the
Gogny energy density functional with two parametrizations, D1S
and D1M. The equal filling approximation is used to describe
odd-$A$ nuclei preserving both axial and time reversal symmetries.
Shape-transition signatures are identified in the $N=60$ isotopes
in both charge radii and spin-parities of the ground states. These
signatures are a common characteristic for nuclei in the whole mass
region. The nuclear deformation and shape coexistence inherent to this
mass region are shown to play a relevant role in the understanding
of the spectroscopic features of the ground and low-lying one-quasiproton
states. Finally, a global picture of the neutron-rich $A\sim100$ mass
region from Krypton up to Molybdenum isotopes is illustrated with
the systematics of the nuclear charge radii isotopic shifts.

\end{abstract}


\maketitle

\section{Introduction}
\label{INTRO}

The structural evolution as a function of the number of nucleons
is a subject of increasing interest in nuclear structure, which
is supported by a very intense activity on both theoretical and 
experimental sides \cite{wood,bender,ours1,ours2,ours3,ours4,
ours5,cheal_2010,lunney,rodriguez,jokinen,mukherjee,charlwood_hi}.
In particular, neutron-rich nuclei in the mass region around
A = 100 have received special attention because of the
interesting nuclear structure features merging there 
\cite{cheal_2010,ours_plb,ours_odd,ours_rb,sarri}.

Experimentally, the efforts are focused on different and
complementary directions. The most relevant for the purpose of
this work are related, on the one hand, to the mass determination 
\cite{lunney,rodriguez,jokinen} in the case of exotic neutron-rich
nuclei with the accuracy required for the modeling of astrophysical
events \cite{apra,goriely}. In particular, the mass region of
our concern here is highly significant to understand the
nucleosynthesis path and the isotopic abundances generated
by the astrophysical r process \cite{cowan}. On the other hand,
the focus is on laser-spectroscopy experiments aimed to measure
nuclear spins, magnetic dipole moments, spectroscopic quadrupole
moments and mean-square charge radii from isotopic shifts.
Considerable progress has been achieved in the last years (for
a recent review, see \cite{cheal_2010} and references therein),
and special attention has received the mass region studied in
this work \cite{charlwood_hi,thibault,buchinger,keim,urban,lhersonneau,
campbell,hager_2006,bucurescu,rahaman,cheal,hager,charlwood,cheal_2009,
hwang,simpson,baczynska}.

Theoretically, the region has been studied using phenomenological
models \cite{skalski,xu02,moller95,moller08} and  microscopic
approaches, based on the Relativistic Mean Field (RMF) \cite{lala},
as well as  nonrelativistic Skyrme \cite{bender06,bender08} and Gogny 
\cite{hilaire,delaroche,ours_plb,ours_odd,ours_rb} energy density
functionals (EDF). All in all, the accumulated information on this
mass region has allowed to establish some characteristic features,
which can be associated to signatures of a shape transition at
$N=60$.

Different nuclear properties sensitive to these structural changes
have been recently investigated \cite{ours_plb,ours_odd,ours_rb}
in several isotopic chains in this mass region. We used a 
selfconsistent Hartree-Fock-Bogoliubov (HFB) approximation based on
the finite range and density dependent Gogny-EDF \cite{gogny} and
the equal filling approximation (EFA) to deal with the odd nucleon. 
We analyzed bulk and spectroscopic properties of neutron-rich 
isotopes  with both even $Z$ (Sr, Zr, and Mo isotopes) and odd $Z$ 
(Rb isotopes). Our purpose in this paper is to complete the
systematic study of the bulk and spectroscopic properties in two
chains of odd-$Z$ isotopes, Yttrium ($Z=39$) and Niobium ($Z=41$),
as well as in the chain of Krypton isotopes ($Z=36$). 

The description of odd-$A$ nuclei involves additional difficulties
because the exact blocking procedure requires the breaking of
time-reversal invariance, making the calculations more involved 
\cite{duguet,bonneau,perez,schunck,exact-bocking-new}. In the
present study we use the EFA, a prescription widely used in
mean-field calculations to preserve the advantages of time-reversal
invariance. The predictions arising from various treatments of the
blocking have been studied in Ref. \cite{schunck}, concluding that
the EFA is sufficiently precise for most practical applications. 
More details of our procedure can be found in Ref. \cite{ours_odd,perez}.

In this work we consider two parametrizations of the Gogny-EDF,
namely D1S \cite{d1s} as the standard and most studied parameter set 
\cite{hilaire,delaroche,egido_04,bertsch,peru} and D1M \cite{d1m}
as the most recent effort to find a new parametrization that improves
the predictions for masses maintaining the excellent performance
and predictive power of the former D1S. Our aim is first to verify
the robustness of our predictions with respect to the particular
version of the EDF employed and second, to test the performance of
D1M in the present context of the spectroscopy of odd-$A$ nuclei.

The paper is organized as follows. In Sec. \ref{THEORY}, we present
a brief description of the theoretical formalism used in the present
work, i.e., the HFB-EFA framework. The results of our calculations
for the considered nuclei are discussed in Sec. \ref{RESULTS}, where
we pay attention to the one-quasiparticle states and their
spectroscopic evolution along the Y and Nb isotopic chains.
We also compare our results with the available experimental data
for charge radii and two-neutron separation energies. 
In Sec. \ref{sec_kr} we show the results for Krypton isotopes, as
well as the systematics of the charge radii in the whole region
from Krypton ($Z=36$) up to Molybdenum ($Z=42$). Finally,
Sec. \ref{CONCLU} is devoted to the concluding remarks and work 
perspectives.


\section{Theoretical framework}
\label{THEORY}

Our theoretical framework to deal with odd-$A$ nuclei is based on
the Gogny-HFB-EFA formalism. In previous studies of even-even
nuclei \cite{ours4,ours5} we have found advantageous
to use the so called gradient method \cite{gradient} to obtain the
solution of the HFB equations, leading to the (even number parity) 
vacuum  $| \Phi \rangle$. Within this method, the HFB equation is 
recast in terms of a minimization (variational) process of the mean
field energy. The Thouless parameters defining the most general 
HFB wave functions \cite{rs} are used as variational parameters. 
As it is customary in calculations with the Gogny force, the 
kinetic energy of the center of mass motion has been subtracted 
from the Routhian to be minimized in order to ensure that the
center of mass is kept at rest. The exchange Coulomb energy was
considered in the Slater approximation and the contribution of
the Coulomb interaction to the pairing field is neglected.
Both axial and time-reversal are selfconsistent symmetries in
our calculations for odd-$A$ nuclei. Triaxial calculations
have also been performed in the case of even-even Kr isotopes. 

The HFB ground-state wave function $ |\Phi \rangle $ of an even-even
nucleus is defined by the condition of being the vacuum of the
annihilation quasiparticle operators $\beta_\mu$ of the Bogoliubov
transformation \cite{rs,mang}. On the other hand,
the ground and low-lying one-quasiparticle states of odd-$A$ systems,
like the ones considered in the present work, can be handled with
blocked (odd number parity \cite{rs,mang}) HFB wave functions

\begin{equation} 
| \Psi_{\mu_B} \rangle  = \beta ^+_{\mu_B} | \Phi \rangle \, ,
\label{one-quasi-state}
\end{equation}
where ${\mu_B}$ indicates the quasiparticle state to be blocked and 
stands for the indexes compatible with the symmetries of the
odd-nuclei, such as the angular momentum projection $K$ and parity
$\pi$ in the case of axial symmetry. As mentioned above, we use here
the EFA to the exact blocking that preserves time-reversal
invariance. In this approximation the unpaired nucleon is treated
on an equal footing with its time-reversed state by sitting half a
nucleon in a given orbital and the other half in its time-reversed
partner. The microscopic justification has been first given in Ref. 
\cite{perez} using ideas of quantum statistical mechanics. The EFA
energy can be obtained as the statistical average, with a given
density matrix operator, and applying the variational principle to
it, the  HFB-EFA equation \cite{perez} is obtained. The
existence of a variational principle allows to use the gradient
method to solve the HFB-EFA equation with the subsequent
simplification in the treatment of the constraints.

The solution of the HFB-EFA (as well as the exact blocked HFB)
equation depends upon the initial blocked level $\mu_B$. In the
HFB-EFA case, the $K$ quantum number is selfconsistently preserved
along the calculation and so does the parity if octupole correlations
are not allowed in the iterative process. Blocking levels with
different $K{^\pi}$ values lead to different quantum states of the
odd-$A$ nucleus, being the ground state the one with the lowest
energy. One should notice that because of the selfconsistent nature
of the whole procedure, for a given $K{^\pi}$, there is no guarantee
that the lowest energy solution for those $K{^\pi}$ values is obtained
by taking the quasiparticle with the lowest energy as the initial
blocked state. Therefore, several quasiparticles with the same 
$K{^\pi}$ must be considered. In addition, in the present case
and because of the presence of coexisting prolate, oblate, and
spherical minima in some of the nuclei considered, blocked
configurations with those quadrupole deformations have to be
explored. 
Constrained calculations have been performed to generate potential
energy curves (PECs) for the even-even neighbor nuclei. One can
find a systematic compilation of PECs obtained with Gogny-D1S in 
Ref. \cite{webpage}. In the case of odd-$A$ nuclei, the purpose
of the computation of such PECs is twofold: First, they give us
initial hints on the evolution of the different competing shapes in
the considered nuclei and, second, they provide a whole set of prolate,
spherical and oblate even-even HFB states (reference states) for the
subsequent treatment of the neighboring odd-$A$ nuclei. 
In fact, once a reference (even-even) state with a
given deformation is chosen, we use it to perform an additional
(constrained) HFB calculation providing an unblocked fully-paired
state corresponding to an odd average neutron number (false vacuum 
\cite{duguet}) with the same deformation. Such  prolate, spherical,
and oblate false vacua are then used as input configurations in our
subsequent blocking scheme (i.e., EFA). 

Thus, the minimization process has to be carried out several times,
using different initial prolate, spherical and oblate (false) vacua.
We have repeated each calculation, for a given false vacuum and $K$
values from 1/2 up to 15/2, using as initial blocking states the 12
quasiparticles corresponding to the lowest quasiparticle energies.
The use of so many initial configurations is to guarantee that we
are not missing the true ground state and all the lowest excited
states. Note  that for nuclei in this region of the nuclear chart,
there exist several competing shapes at low excitation energy and
therefore our procedure assures that the lowest energy solution can
be reached for all values of the quadrupole moment $Q_{20}$ and mass
number.
   

\section{Results for odd-A Yttrium and Niobium isotopes}
\label{RESULTS}

Being odd-$Z$ nuclei, the spin and parity of odd-$A$ Yttrium and
Niobium isotopes are determined by the state occupied by the unpaired
proton. The spectroscopic properties of the odd-$A$ isotopes are
determined by the one-quasiproton configurations that, in principle,
are expected to be rather stable against variations in the number
of even neutrons. However, as it is known in neighboring nuclei,
approaching $N\sim 60$, the isotopes become well deformed 
\cite{wood,ours_plb} and the abrupt change in deformation induces
signatures in nuclear bulk properties like the two-neutron separation
energies and the nuclear charge radii, as well as in spectroscopic
properties. In particular, the spin and parity of the nuclear ground
state might flip suddenly from one isotope to another, reflecting 
the structural change.

\subsection{Low-lying one-quasiparticle states}

\begin{figure*}[h]
\centering
\includegraphics[width=150mm]{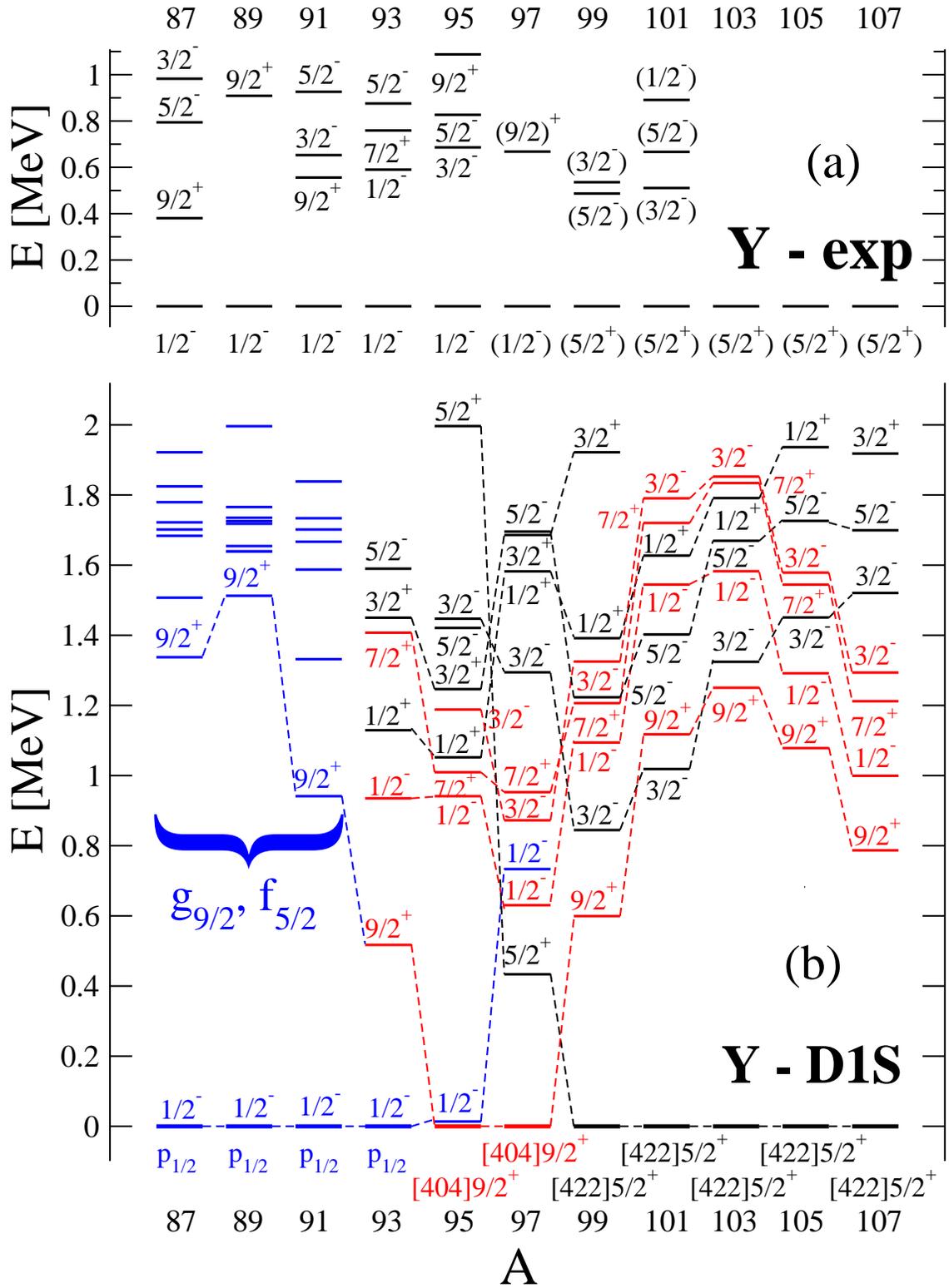}
\caption{
(Color online) Experimental (a) excitation energies and spin-parity 
assignments of the non-collective states are compared with  HFB-EFA
results (b) for the  one-quasiproton states in odd-$A$ Yttrium isotopes.
Prolate configurations are shown by black lines, oblate ones by red lines,
and spherical ones by blue lines.}
\label{fig_y}
\end{figure*}

In Fig. \ref{fig_y} we can see the experimental excitation energies
and spin-parity assignments (a) in odd-$A$ neutron-rich Yttrium
isotopes \cite{exp_ensdf}. They are compared to the one-quasiproton
states predicted by our Gogny-D1S HFB-EFA calculation (b), where the
excited states in a given isotope are referred to the corresponding
ground state, regardless its shape. Prolate configurations in our
calculations are shown by black lines, oblate ones by red lines, and
spherical ones by blue lines. The quasiparticle states are labeled
by their $K^\pi$ quantum numbers. The most important configurations
are joined by dashed lines following the isotopic evolution. In
addition, the ground states are labeled by their asymptotic quantum
numbers $[N,n_z,\Lambda]K^\pi$.

Experimentally, we observe $J^\pi=1/2^-$ ground states in 
$^{87,89,91,93,95,97}$Y, then we have $(5/2^+)$ states in the heavier
isotopes $^{99,101,103,105,107}$Y, although their assignments are in 
these cases uncertain. The $9/2^+$ states appear as excited states in
the lighter isotopes, while $3/2^-$ and $5/2^-$ excited states 
are also found in most isotopes, but especially in the heavier ones.
The most striking feature observed is the abrupt
transition at $N=60\, (A=99)$ from $1/2^-$ to $5/2^+$ ground states.

The theoretical interpretation of these findings can be understood
from the analysis of our results in the lower figure (b). According
to our calculations, Yttrium isotopes evolve from spherical shapes
in $^{87-93}$Y around $N=50$, with the spherical $p_{1/2}$ shell in
the ground state and close $g_{9/2}$ and $f_{5/2}$ as excited states,
to slightly deformed oblate shapes in $^{95,97}$Y, and finally to
well deformed prolate shapes in $^{99-107}$Y. 
In the lighter isotopes the ground states correspond to $1/2^-$
states ($p_{1/2}$), whereas the excited states correspond to the split
$K^\pi$ levels coming from $g_{9/2}$ and $f_{5/2}$, with $9/2^+$ states
as the lowest excited states in agreement with the measured low-lying
spectra. The excited $9/2^+$ states decrease in energy as we move
away from the magic neutron number $N=50$ because of the incipient
oblate deformation emerging, and they eventually become the ground
state in $^{97}$Y. In the case of the $^{95}$Y isotope the oblate
$9/2^+$ state is practically degenerate with the $1/2^-$ spherical
state. In our description the 
nucleus $^{97}$Y displays a $9/2^+$ oblate ground state with a $1/2^-$ 
excited state at 0.7 MeV and an incipient prolate $5/2^+$ at 0.4 MeV
that will become the ground state in the heavier isotopes $^{99-107}$Y.
All of these heavier prolate isotopes present oblate $9/2^+$ excited
states at energies in the range $0.6-1.2$ MeV. Thus, the experimentally
observed leap of the ground-state spin-parity assignments at $N=60$
is well accounted for by the present calculations, which is interpreted
as a sudden ground state shape change.

Similarly to Fig. \ref{fig_y}, we show in Fig. \ref{fig_nb} the 
corresponding results for Niobium isotopes. Experimentally 
\cite{exp_ensdf} we observe $J^\pi=9/2^+$ ground states in the lighter
$^{89-99}$Nb isotopes with $1/2^-$ excited states at close energy and
then we find $(5/2^+)$ ground states in the heavier isotopes 
$^{101-105}$Nb, although as in the case of  Y  isotopes, the spin-parity
assignments are still uncertain in these isotopes. A characteristic
jump at $N=60\, (A=101)$ from $9/2^+$ to $(5/2^+)$ ground states is
once more found. The calculations from Gogny-D1S (b) show in the
lighter isotopes a clear correspondence between the measured ground
state $J^\pi=9/2^+$ and the calculated ones. The observed excited
states $1/2^-$ are also labeled in the calculations. They correspond
to the excited states from the $p_{1/2}$ shell. However, at variance
with experiment, our results indicate a transition in $^{99}$Nb
to the $7/2^+$ oblate state. This oblate configuration is kept all
the way up to the heaviest isotope considered $^{109}$Nb. The
observed $(5/2^+)$ ground states are found in our calculations as
prolate configurations at excitation energies between 0.6 and 
1 MeV in $^{101,103,105}$Nb. It will be interesting to see whether 
the experimental ground states of $^{101,103,105}$Nb are confirmed to
be $5/2^+$ states and to measure the heavier $^{107,109}$Nb, where 
according to the calculations the $5/2^+$ states appear at a very
high excitation energy relative to the ground state.

\begin{figure*}[h]
\centering
\includegraphics[width=150mm]{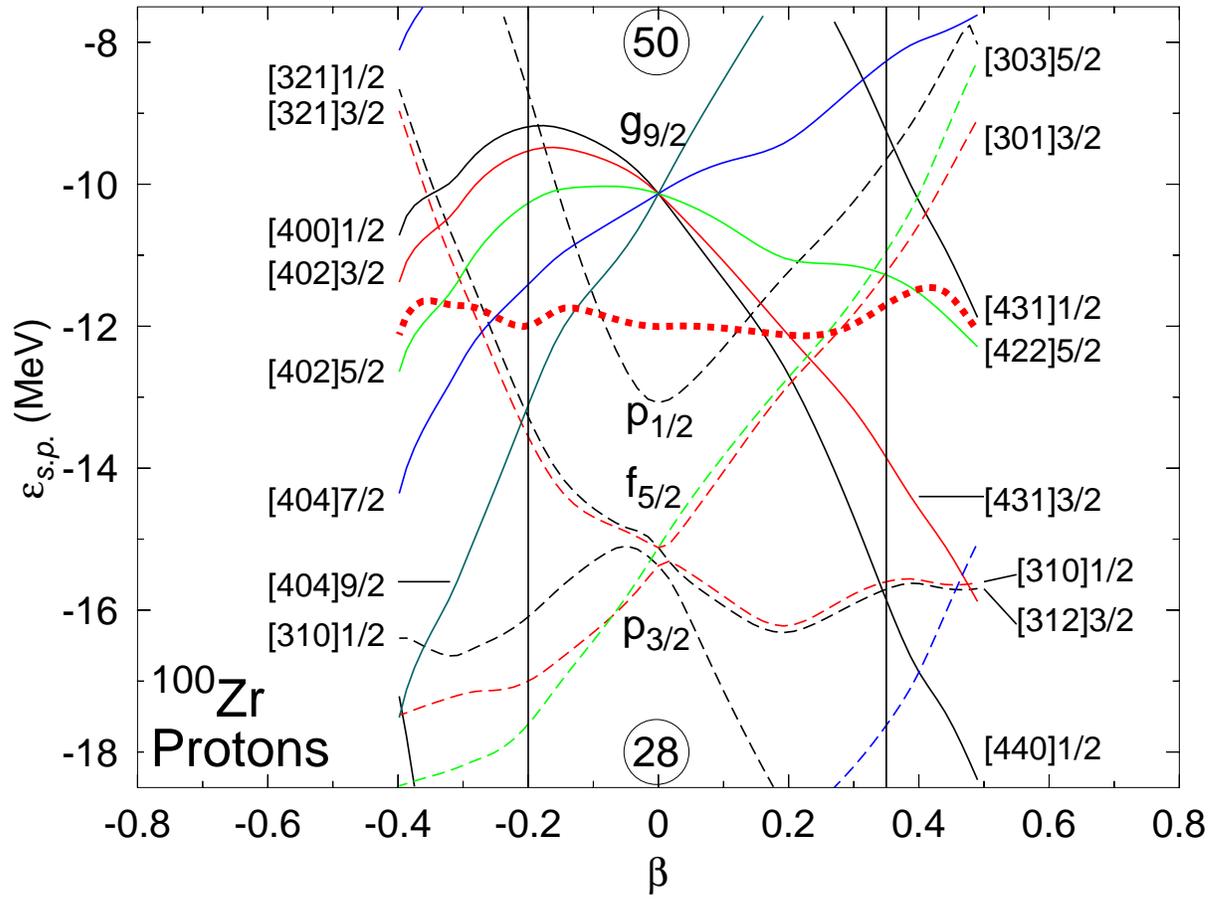}
\caption{(Color online) Proton single particle energies for $^{100}$Zr as a
function of the quadrupole deformation parameter $\beta$. The Fermi level
is plotted by a thick dotted line. The results correspond to the Gogny-D1S
EDF. Solid lines correspond to levels with positive parity, whereas dashed
lines correspond to negative parity states. Asymptotic Nilsson quantum
numbers $[N,n_z,\Lambda]K^\pi$ are also shown at the vertical lines where
the minima of the potential energy curves are located.}
\label{fig_spe}
\end{figure*}

The disagreement between the theoretical predictions and experimental
data for the spin and parity of the heavier Nb isotopes could
be understood if triaxiality effects are invoked. In Fig. \ref{fig_spe},
where the proton single particle levels for $^{100}$Zr are depicted
as functions of the deformation parameter $\beta$ \cite{ours2}, we
observe for $\beta=-0.2$ (the position of the oblate minimum) a 
$K^{\pi}=7/2^+$ orbital just above the Fermi level. The occupancy of
this orbital produces the $7/2^+$ oblate ground state in $^{101}$Nb.
The $K^{\pi}=5/2^+$ orbital coming from the same $g_{9/2}$ subshell and
presumably responsible for the experimental spin and parity lies
higher in energy. The situation is reversed in the prolate side where
at $\beta=0.35$ the $K^{\pi}=5/2^+$ is below the $K^{\pi}=7/2^+$ and is
still above the Fermi level. The easiest way to connect the oblate
and prolate sides is through the triaxial $\gamma$ degree of freedom, 
with $\gamma$ ranging from $60^\circ$ (oblate side) to $0^\circ$ in
the prolate side. It is obvious that the energy of the $K^{\pi}=5/2^+$
orbital has to go below the energy of the $K^{\pi}=7/2^+$ one at some
$\gamma$ value and it is very likely that at that point the blocking
of the $K^{\pi}=5/2^+$ orbital will produce the ground state minimum.
This is obviously a hand-waving argument as the $K$ quantum number is
not preserved along the $\gamma$ path but again, it is very likely,
that both components $K^{\pi}=5/2^+$ and $K^{\pi}=7/2^+$ are going to
be dominant in the orbital just above the Fermi level. The above
argument calls for the necessity of carrying out a full triaxial
calculation for the heaviest odd-Z Nb isotopes which is, for the
moment, not possible as it will require an extension of our present
computational codes for odd nuclei to include the role of  
triaxiality. Work along these lines is in progress.

\begin{figure*}[h]
\centering
\includegraphics[width=150mm]{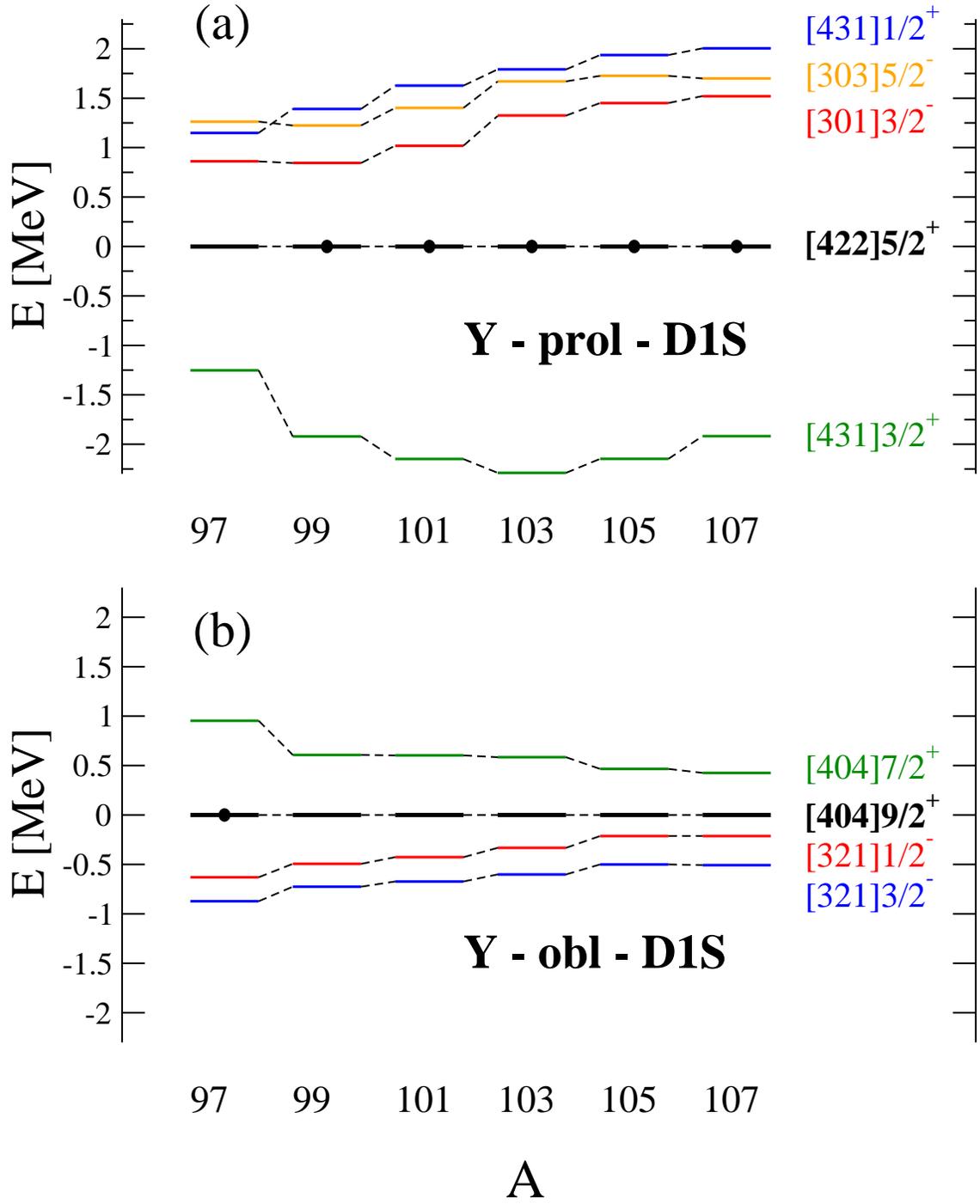}
\caption{(Color online) Gogny-D1S excitation energies of single-quasiproton
prolate (a) and oblate (b) states in Yttrium isotopes. Hole states are plotted
below zero energy and particle states are plotted above. The absolute
ground states are indicated with a circle.}
\label{fig_y_def_d1s}
\end{figure*}

\begin{figure*}[h]
\centering
\includegraphics[width=150mm]{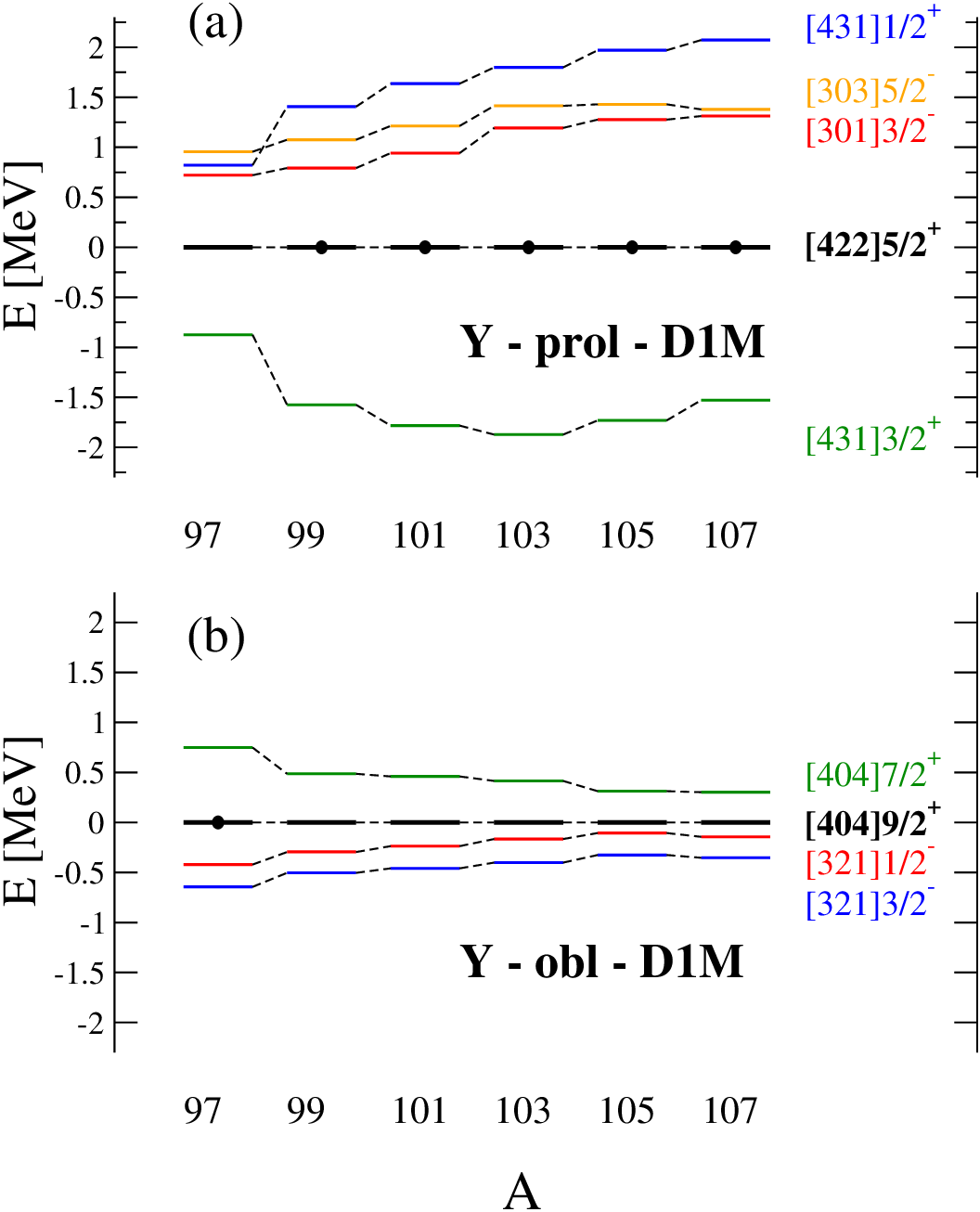}
\caption{(Color online) Same as in Fig. \ref{fig_y_def_d1s}, but for Gogny-D1M.}
\label{fig_y_def_d1m}
\end{figure*}

\begin{figure*}[h]
\centering
\includegraphics[width=150mm]{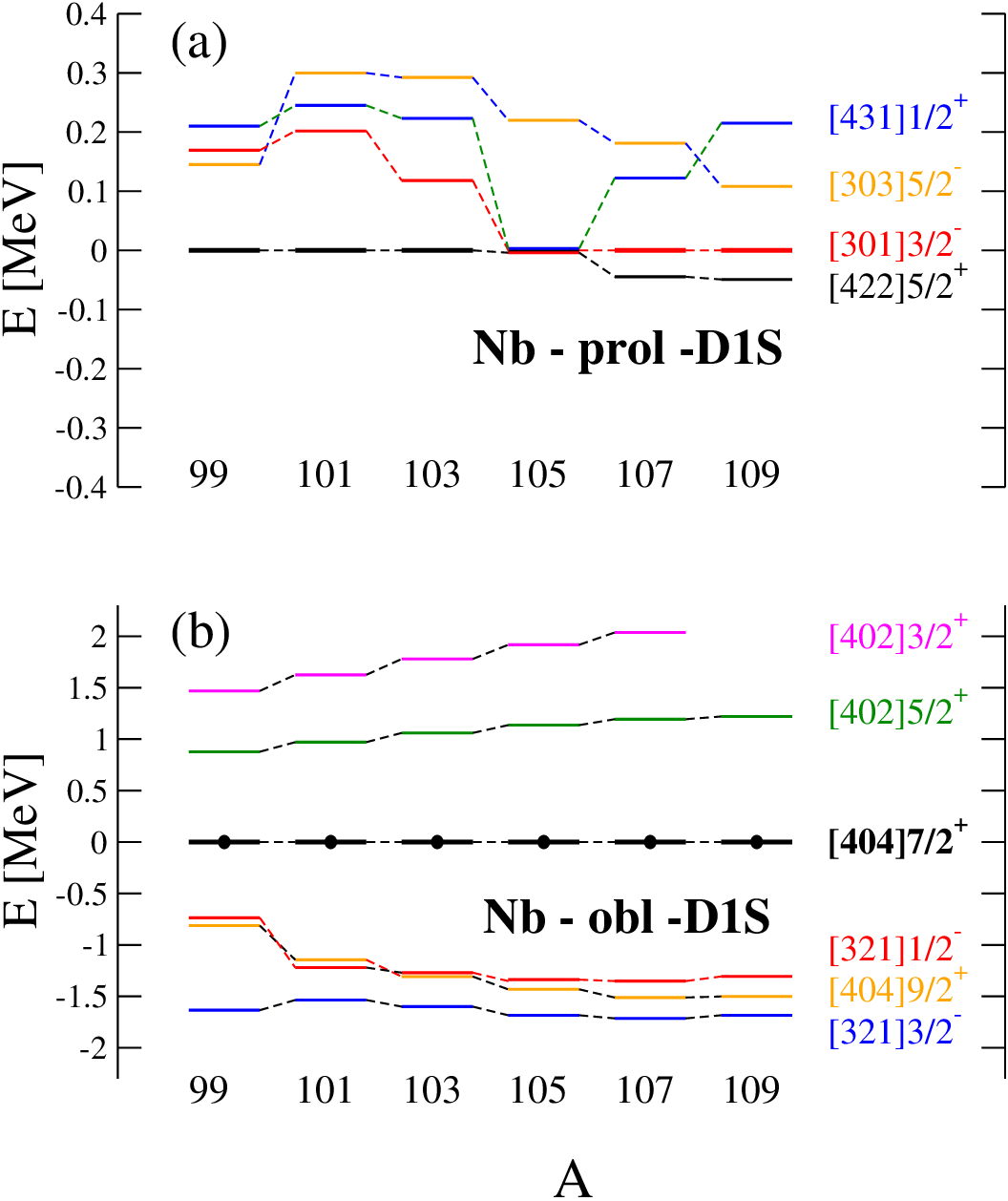}
\caption{(Color online) Same as in Fig. \ref{fig_y_def_d1s}, but for Niobium
isotopes.}
\label{fig_nb_def_d1s}
\end{figure*}

\begin{figure*}[h]
\centering
\includegraphics[width=150mm]{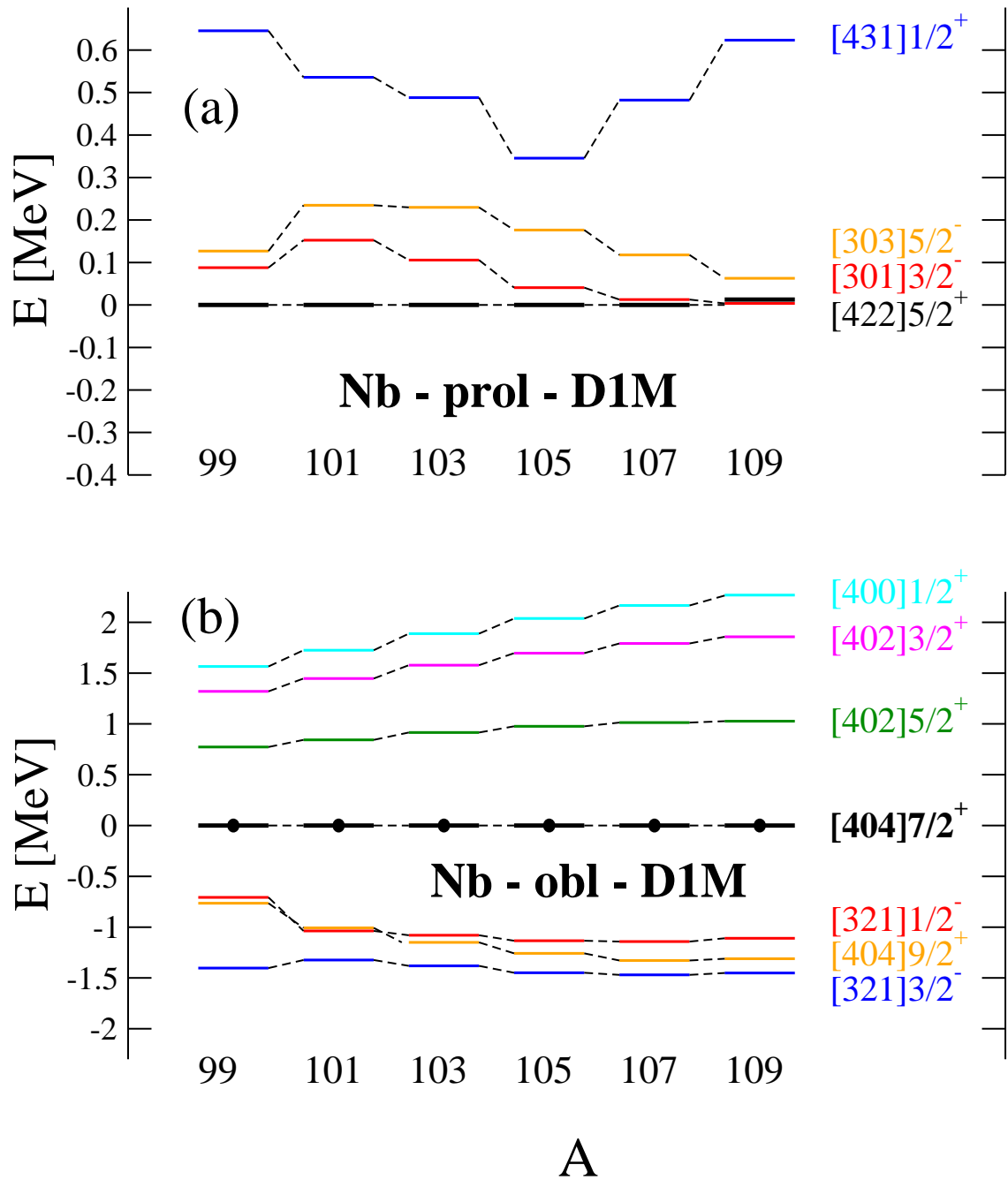}
\caption{(Color online) Same as in Fig. \ref{fig_y_def_d1m}, but for
Niobium isotopes.}
\label{fig_nb_def_d1m}
\end{figure*}

In the next figures we compare the spectroscopic properties of the
Gogny D1S and D1M parametrizations. In Figs. \ref{fig_y_def_d1s}, 
\ref{fig_y_def_d1m}, \ref{fig_nb_def_d1s}, and \ref{fig_nb_def_d1m}
we have separated the prolate (a) and oblate (b) states and have
plotted the most relevant hole states below zero energy and the
particle states above. The ground states, either oblate or prolate,
are indicated with a circle.
Specifically, Figs. \ref{fig_y_def_d1s} and \ref{fig_y_def_d1m} 
correspond to Y isotopes with Gogny-D1S and Gogny-D1M, respectively,
whereas Figs. \ref{fig_nb_def_d1s} and \ref{fig_nb_def_d1m} 
correspond to Nb isotopes with Gogny-D1S and Gogny-D1M, respectively.

In Figs. \ref{fig_y_def_d1s} and \ref{fig_y_def_d1m} for Yttrium,
in the prolate case (a) we have depicted the evolution of the 
$5/2^+(g_{9/2})$ states, which are ground states for $^{99-107}$Y. 
In the upper region we find the $3/2^-(f_{5/2})$, $5/2^-(f_{5/2})$, 
and $1/2^+(d_{5/2})$, while in the lower region we have the 
$3/2^+(g_{9/2})$, as it can be seen from Fig. \ref{fig_spe} on the 
prolate side. Similar states are found in the prolate graphs (a) 
of Figs. \ref{fig_nb_def_d1s} and \ref{fig_nb_def_d1m} for Nb isotopes,
but in this case the prolate configurations are not ground states.
In the oblate case (b) we can see for Y isotopes the $9/2^+(g_{9/2})$
state, which is ground state in $^{97}$Y, the $7/2^+(g_{9/2})$ as a 
particle state and the $1/2^-$ and $3/2^-$ from $f_{5/2}$
as hole states, as it can be seen from Fig. \ref{fig_spe} on the 
oblate side. The oblate configurations in Nb isotopes show the 
$7/2^+(g_{9/2})$ as the ground state in all the isotopes depicted.
In addition to the states shown for Y isotopes, we also show here
the particle states $5/2^+$ and $3/2^+$ from $g_{9/2}$.

Very similar results are obtained from both parametrizations
of the Gogny-EDF. The only difference worth mentioning is that
D1M produces slightly lower excited states and then a more compact
level density. This feature can be understood from its larger
effective mass that makes the single particle spectrum somewhat more
dense with D1M. This answers our original question about robustness
of the calculations and the reliability of D1M in what concerns the
spectroscopic properties of the two parametrizations.

\begin{figure*}[h]
\centering
\includegraphics[width=150mm]{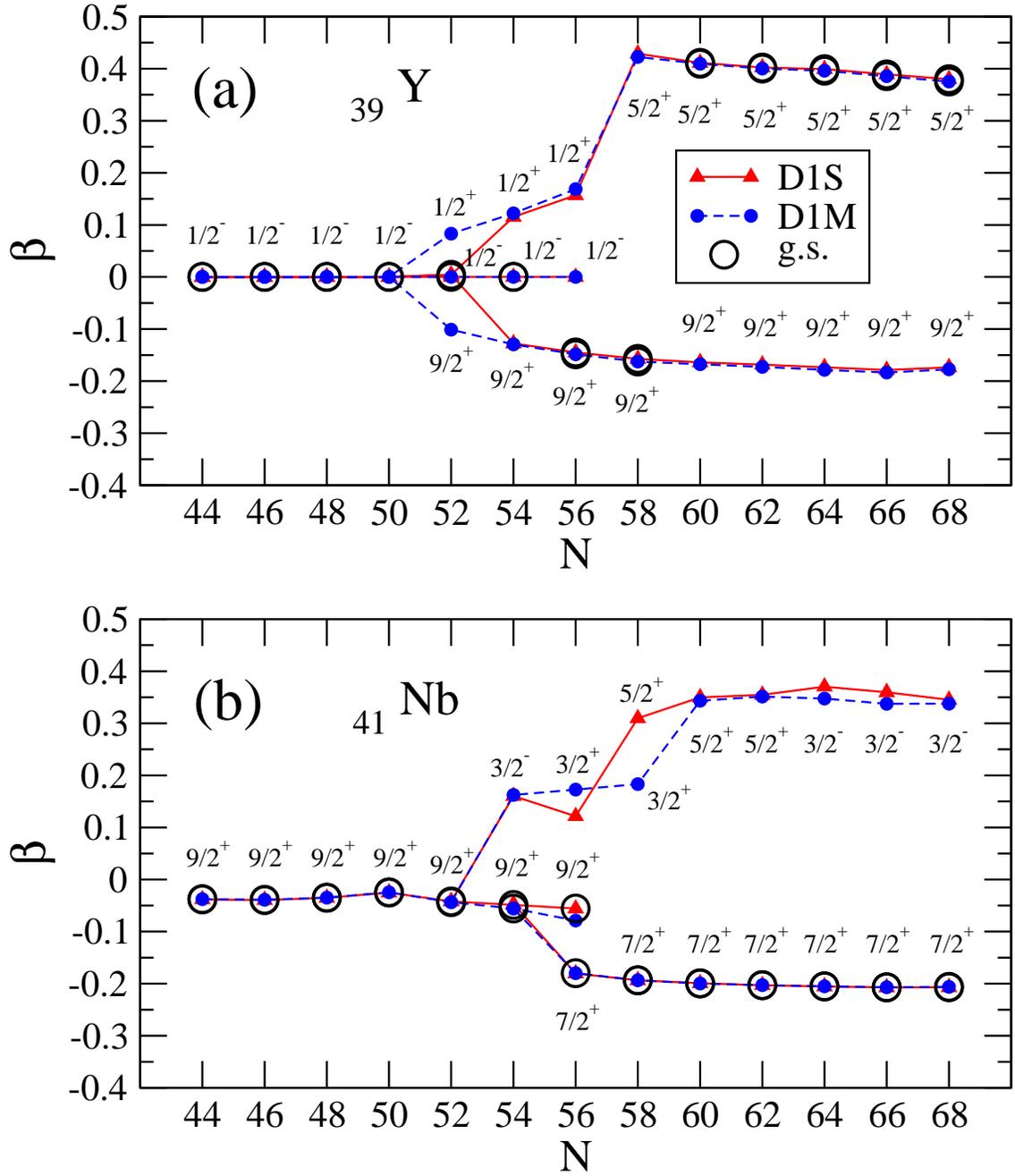}
\caption{(Color online) Isotopic evolution of the quadrupole deformation
parameter $\beta$ of the energy minima obtained from Gogny D1S and D1M
calculations for Yttrium (a) and Niobium (b) isotopes.}
\label{fig_beta}
\end{figure*}

To further illustrate the role of deformation and spin-parity
assignments in the isotopic evolution, we display in Fig. \ref{fig_beta}
for Y (a) and Nb (b) isotopes, the axial quadrupole deformation 
$\beta$ of the energy minima as a
function of $N$ with both D1S and D1M parametrizations. The deformation
of the ground state for each isotope is encircled. We can see that
beyond the semi-magic isotope with $N=50$ we start getting two minima
in the prolate and oblate sectors. In the case of Y isotopes the spherical
$1/2^-$ states are ground states up to $N=54$ in agreement with experiment.
For $N=52$ and $N=54$ one can see the appearance of an oblate $9/2^+$ and
a prolate $1/2^+$ excited states, which are almost degenerate with the
ground state. The next isotopes, with $N=56$ and $N=58$, display a slightly
oblate  $9/2^+$ ground state and then suddenly a transition occurs
at $N=60$ to well deformed prolate $5/2^+$ ground states, in 
agreement with experiment. The calculations with both D1S and D1M are
coincident. On the other hand, in the case of Nb isotopes, we observe
the spherical $9/2^+$ states
being ground states along $N=44-56$ in agreement with experiment.
However, starting at $N=56$ in the case of D1M and at $N=58$ in the
case of D1S, oblate $7/2^+$ ground states are obtained
at variance with
the experimental assignment ($5/2^+$) for these states, which is 
nevertheless still uncertain, as already mentioned above.

A qualitative understanding of the features just discussed can be
obtained by looking at the single particle levels for protons in
$^{100}$Zr and depicted in Fig. \ref{fig_spe} as a function of the
beta deformation parameter. This $Z=40$ even-even nucleus is in between
Y and Nb regarding the number of protons and therefore its single
particle properties should not differ too much (at least at a qualitative
level) from the ones of Y and Nb. The insight obtained from Fig. \ref{fig_spe}
can only be qualitative owing to the selfconsistent character of the EFA
that somehow incorporates the polarization effects induced by the single
proton in the even-even core. 

\subsection{Charge radii and two-neutron separation energies}

\begin{figure*}[h]
\centering
\includegraphics[width=150mm]{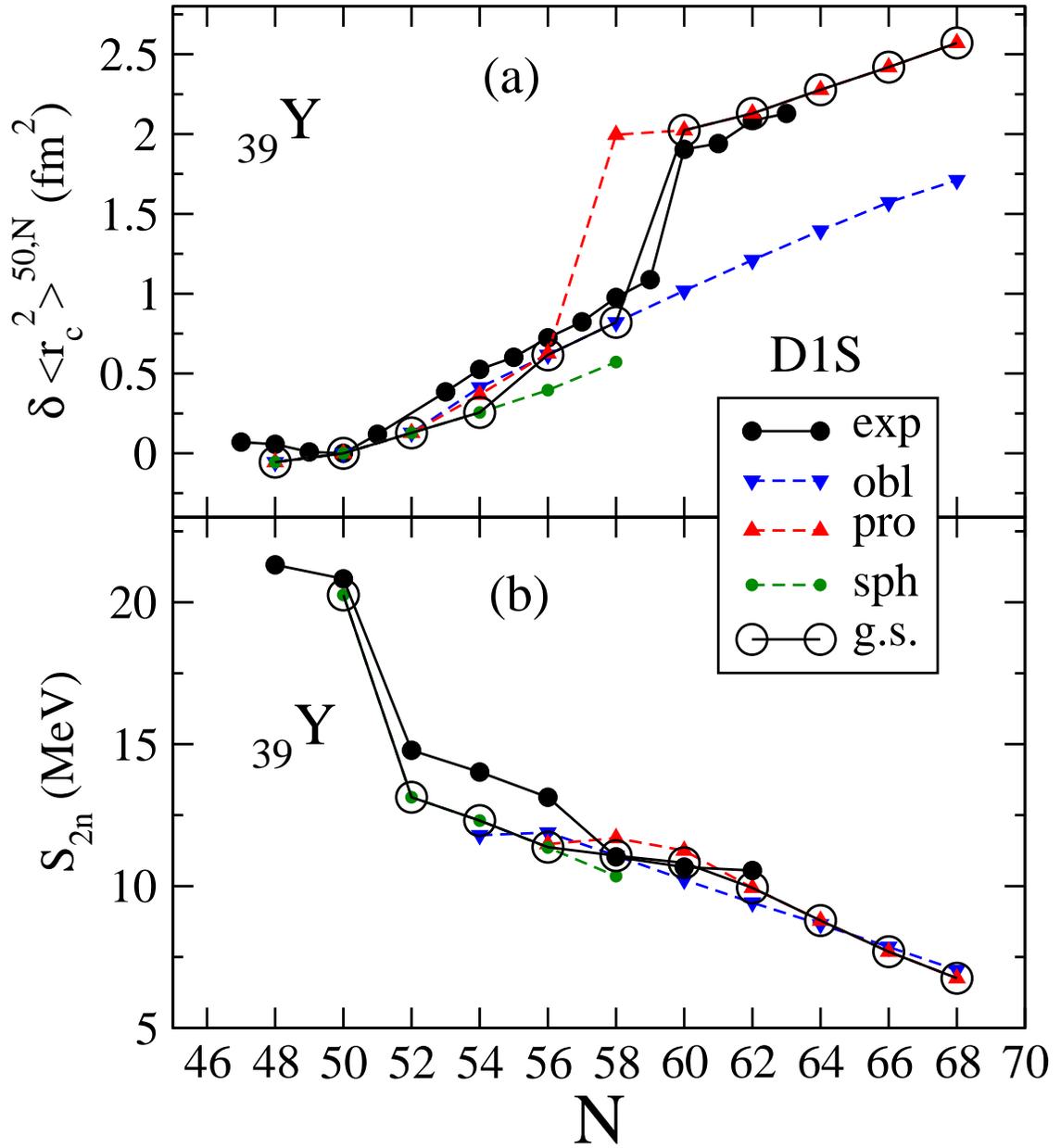}
\caption{(Color online) Gogny-D1S HFB results for $\delta \langle r^2_c \rangle$ 
(a) and $S_{2n}$ (b) in odd-$A$ Yttrium isotopes compared to experimental data
from Ref. \cite{audi,hager} for masses and from Ref. \cite{baczynska,cheal}
for radii. Results for prolate, oblate, and spherical minima are displayed
with different symbols (see legend). Open circles correspond to ground-state
results.}
\label{fig_s2n_d1s_y}
\end{figure*}

\begin{figure*}[h]
\centering
\includegraphics[width=150mm]{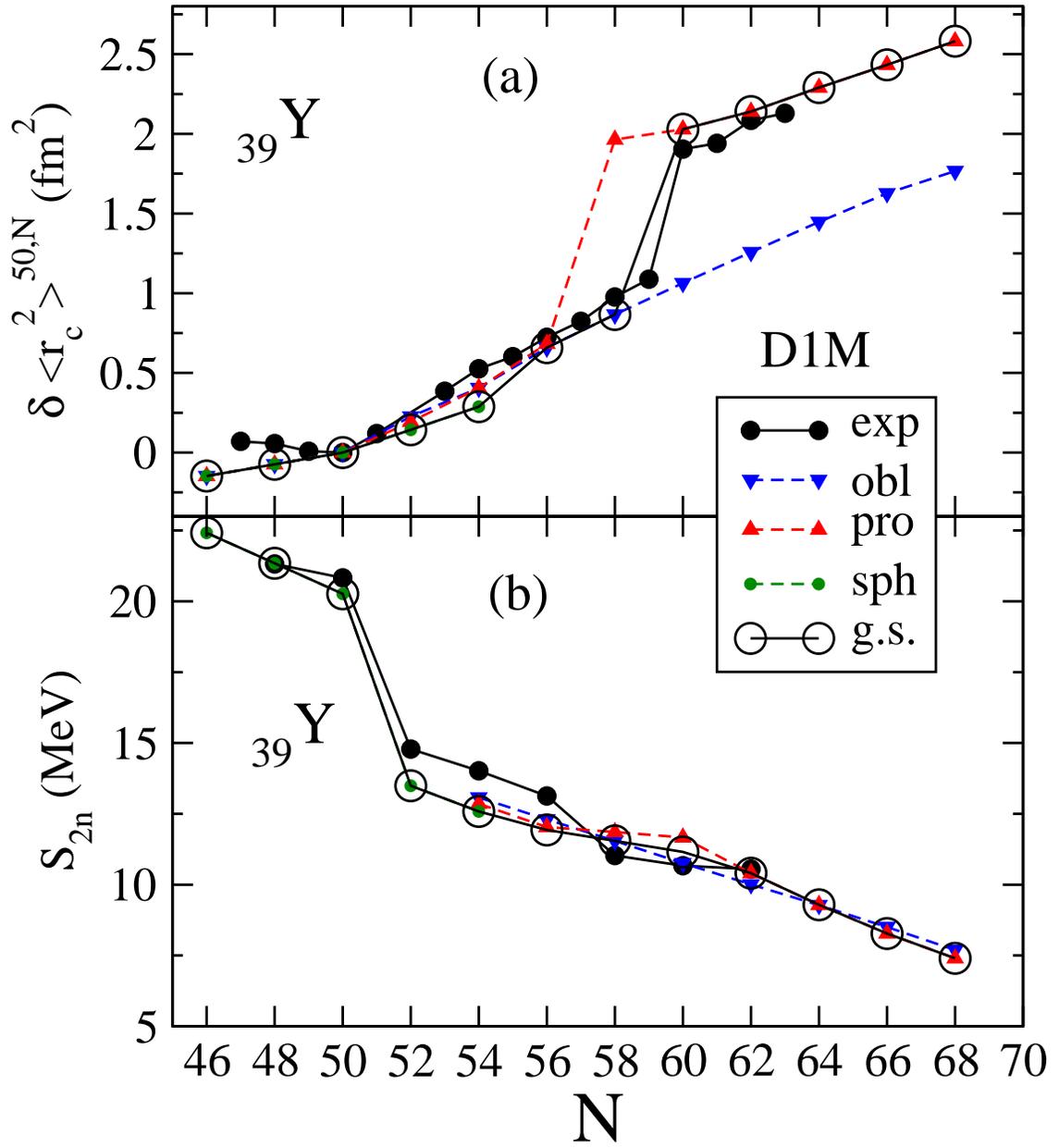}
\caption{(Color online) Same as in Fig. \ref{fig_s2n_d1s_y}, but for Gogny-D1M.}
\label{fig_s2n_d1m_y}
\end{figure*}

\begin{figure*}[h]
\centering
\includegraphics[width=150mm]{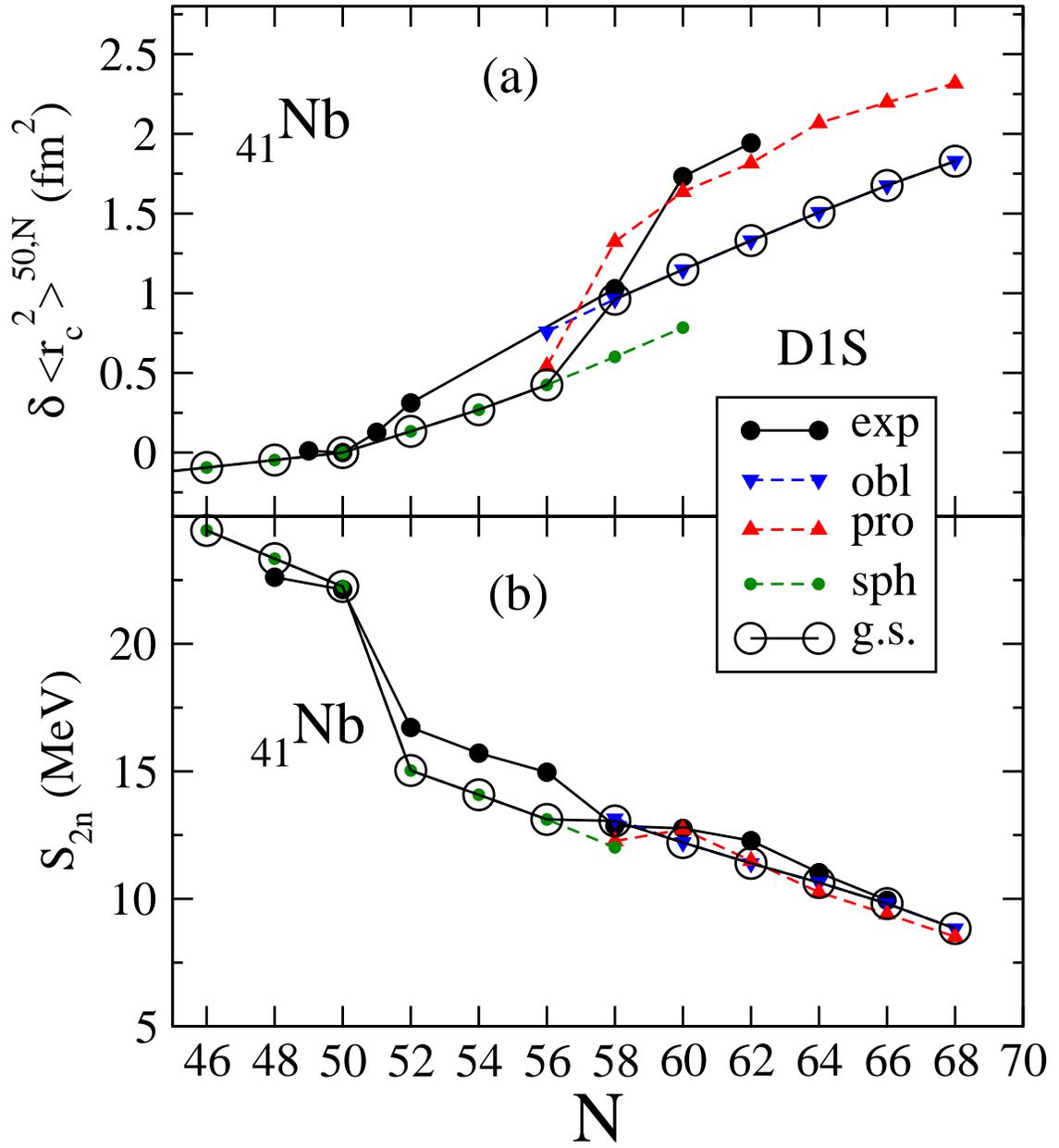}
\caption{(Color online) Same as in Fig. \ref{fig_s2n_d1s_y}, but for Niobium
isotopes. Experimental data for radii are from Ref. \cite{cheal_2009} }
\label{fig_s2n_d1s_nb}
\end{figure*}

\begin{figure*}[h]
\centering
\includegraphics[width=150mm]{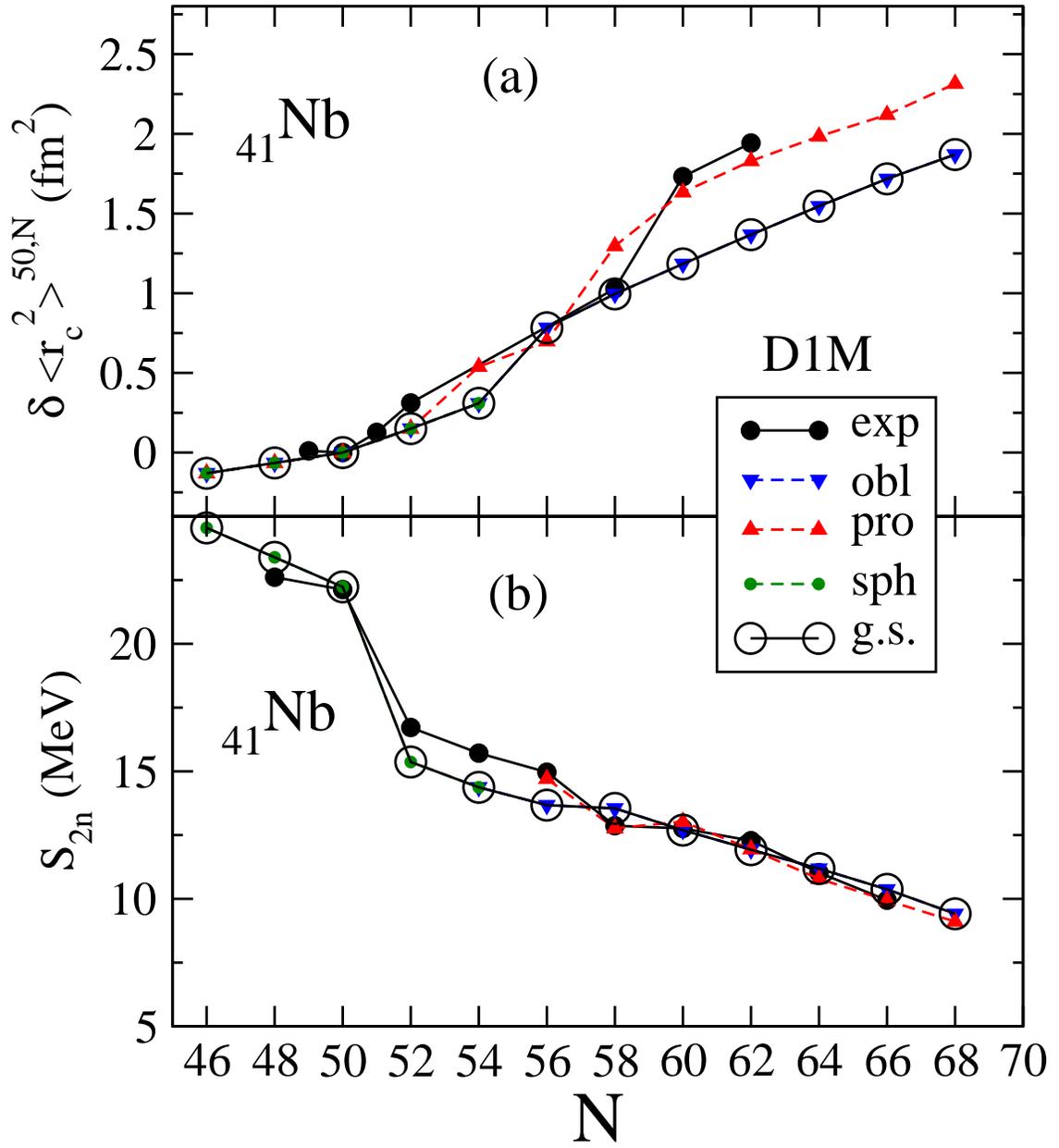}
\caption{(Color online) Same as in Fig. \ref{fig_s2n_d1s_y}, but for Niobium
isotopes and Gogny-D1M.}
\label{fig_s2n_d1m_nb}
\end{figure*}

In  Figs. \ref{fig_s2n_d1s_y}, \ref{fig_s2n_d1m_y},
\ref{fig_s2n_d1s_nb}, and \ref{fig_s2n_d1m_nb}, we show the results
for the charge radii differences (a) defined as 
$\delta \langle r^2_c \rangle ^{50,N}= \langle r^2_c 
\rangle ^N - \langle r^2_c \rangle ^{50}$, calculated with the same
corrections as in Ref. \cite{ours_plb}, and for the two-neutron
separation energies $S_{2n}$ (b).
Figures \ref{fig_s2n_d1s_y} and \ref{fig_s2n_d1m_y} show the results
for Yttrium isotopes calculated with Gogny-D1S and Gogny-D1M, respectively.
Figures \ref{fig_s2n_d1s_nb} and \ref{fig_s2n_d1m_nb} are the
corresponding ones for Niobium isotopes. 
Our results are compared with isotope shifts from laser spectroscopy
experiments \cite{baczynska,cheal,cheal_2009} in the case of 
$\delta \langle r^2_c \rangle $ and with mass measurements from Refs. 
\cite{audi,hager} in the case of $S_{2n}$.
We plot the results corresponding to  the spherical, oblate, and prolate
shapes at the  minima of the PECs. Results corresponding to the ground
states are encircled in each isotope.

The first thing to notice is the remarkable similarity between the
results obtained with Gogny-D1S and Gogny-D1M. The only visible
difference between D1S and D1M is a somewhat better agreement with
the experimental $S_{2n}$ values in the case of D1M, as it can be 
expected from its improved fitting protocol, focusing in
a more accurate reproduction of the nuclear masses

For Y isotopes, the measured nuclear charge radii differences depicted
in (a) exhibit a sizable jump at $N=60$ where the radius suddenly
increases. This observation is well accounted for in our calculations,
where the encircled ground states show that a jump of the same
magnitude occurs between $N=58$ and $N=60$. As we can see, the
origin of the jump is related to the shape transition from slightly
oblate shapes to well deformed prolate configurations beyond $N=60$.
The charge radius demonstrates once more to be a nuclear property
very sensitive to those shape variations. It is also worth mentioning
that the sudden change of the radius is perfectly correlated with the
change in the spin-parity of the ground states studied above.

In the case of Nb isotopes, shown  in Figs. \ref{fig_s2n_d1s_nb} and 
\ref{fig_s2n_d1m_nb}, the data are still scarce \cite{cheal_2009}, 
but nevertheless a jump between $N=58$ and $N=60$ is also experimentally
observed, although not as sharp as in the previous case. This is also
associated with the transition from $9/2^+$ to $(5/2^+)$ observed in
these isotopes between $N=58$ and $N=60$. As we have mentioned, the
calculations produce only a gradual transition from spherical to oblate
shapes, but the prolate shapes associated to the $5/2^+$ states never
become ground states. In order to follow the experimental trend we will
need a transition at $N=60$ to get our calculations to the data points,
which are explained in any case by the prolate shapes.
We faced in the past \cite{ours_plb} similar problems in the case of
Mo ($Z=42$) isotopes, which are the neighbors of Nb ($Z=41$). In that
case it was  demonstrated that, for even-even Mo isotopes, the 
incipient emergence of triaxiality was at the origin of the experimentally
observed radius evolution. The quadrupole deformations of the triaxial
minima were in Mo isotopes very close to the prolate ones and therefore,
their effect on the radii where also similar to the prolate case.
Similar type of arguments could be use here for Nb isotopes and one
would expect the triaxial degrees of freedom to be more involved, but
our present technical capability prevents us to carry out triaxial
calculations for odd-$A$ nuclei.

The results for $S_{2n}$ agree in general with the measurements, but
we get systematically a shell effect at $N=50$ stronger than experiment.
This is a well known feature of any mean field approach \cite{bender08},
which is cured once beyond mean-field configuration mixing calculations,
in the spirit of the Generator Coordinate Method (GCM), are implemented
in the formalism. In any case, it would be very helpful to extend and
improve mass measurements, reducing the still large uncertainties in
neutron-rich isotopes.  Most of the reported measurements in this
exotic mass region are based on $\beta$-endpoint measurements, which
are not completely reliable because they lead to very strong binding
\cite{hager}.

\section{Charge radii systematics in the Kr-Mo region}
\label{sec_kr}

In this section we discuss the isotopic evolution of the charge radii
in the whole region from Kr ($Z=36$) up to Mo ($Z=42$). These isotope 
shifts have been found to be very sensitive probes of nuclear shape
transitions and it is worth studying globally the whole Kr-Mo region
discussing the similarities and differences among the various isotopic 
chains.

\begin{figure*}[h]
\centering
\includegraphics[width=150mm]{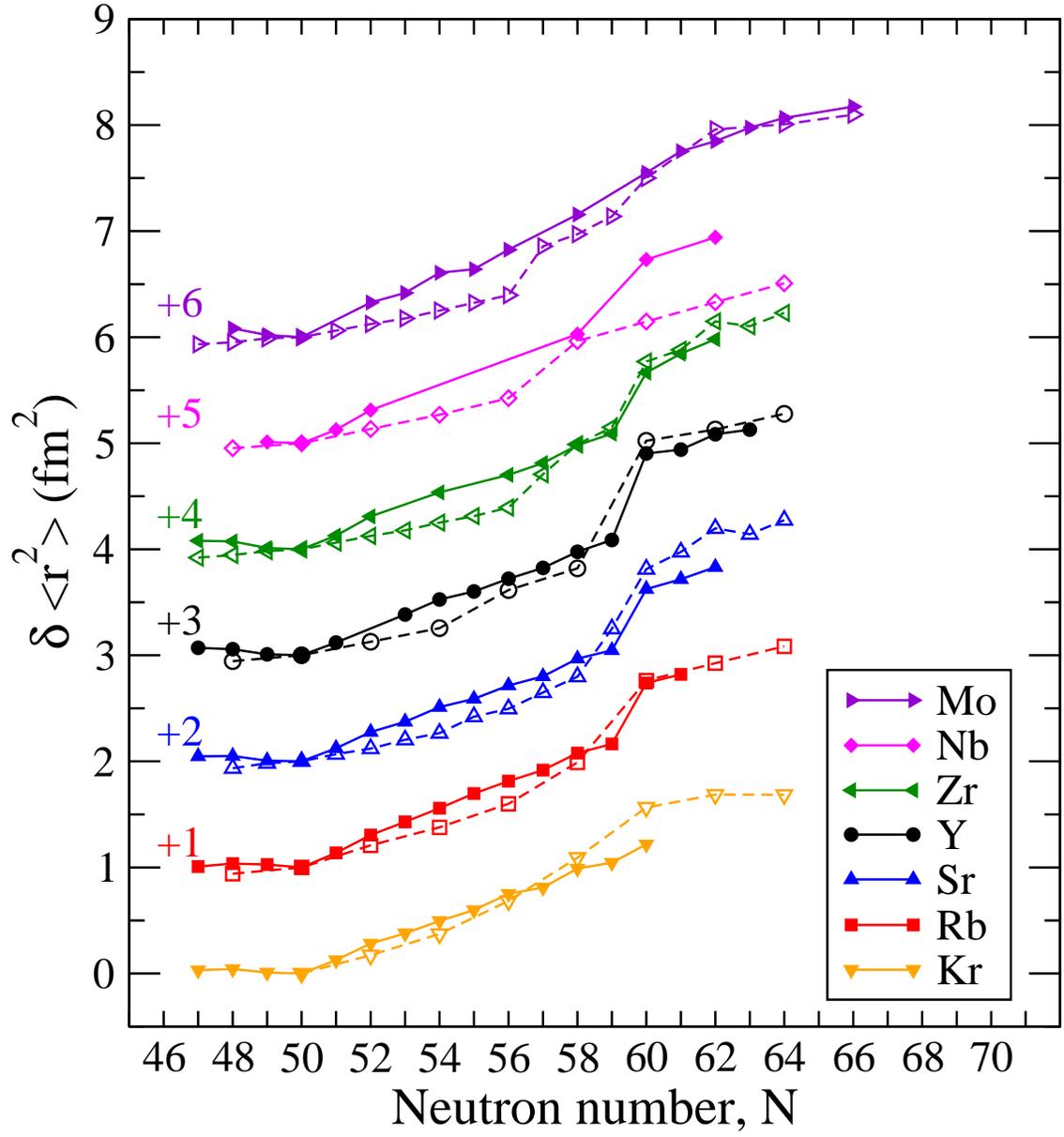}
\caption{(Color online) Gogny-D1S HFB results for $\delta \langle r^2_c \rangle$ 
compared to the measured values for Kr, Rb, Sr, Y, Zr, Nb, and Mo
isotopic chains.}
\label{fig_dr2_global}
\end{figure*}

In Fig. \ref{fig_dr2_global} we have compiled the measured isotope
shifts $\delta \langle r^2_c \rangle $ for Mo \cite{charlwood},
Nb \cite{cheal_2009}, Zr \cite{campbell}, Y \cite{baczynska,cheal}, 
Sr \cite{buchinger}, Rb \cite{thibault}, and Kr \cite{keim} isotopes.
These data are shown with solid symbols and connected with continuous 
lines. In addition, we can also find our results from Gogny-D1S-HFB
calculations, which are shown with open symbols connected with dashed
lines. Very similar results are provided by the Gogny-D1M EDF.
As we can see in Fig. \ref{fig_dr2_global}, this mass region is
characterized by a jump of the mean square charge radii at around
$N=60$, which reflects the sudden increase of deformation or shape
transition that occurs for these isotones. The increase in the charge
radii is maximum for Y isotopes in the middle region and it corresponds
to a change from oblate to prolate shapes at $N=60$, as it was shown
in the previous section. 
Around Y nuclei, we find that Zr and Sr isotopes also show this big
effect that was studied in Refs. \cite{ours_plb,ours_odd} and interpreted
as an oblate-prolate transition as well. The same is also true for Rb
isotopes, as discussed in Ref. \cite{ours_rb}. The behavior of Mo isotopes
was studied in Refs.  \cite{ours_plb,ours_odd}, where it was shown that
the suppression of the jump has its origin in the onset of triaxiality
that smooths the evolution of the radii. The case of Nb isotopes studied
in this paper shows strong similarities with the case of Mo isotopes and
triaxiality is likely again the reason for the discrepancies. 

\begin{figure*}[h]
\centering
\includegraphics[width=150mm]{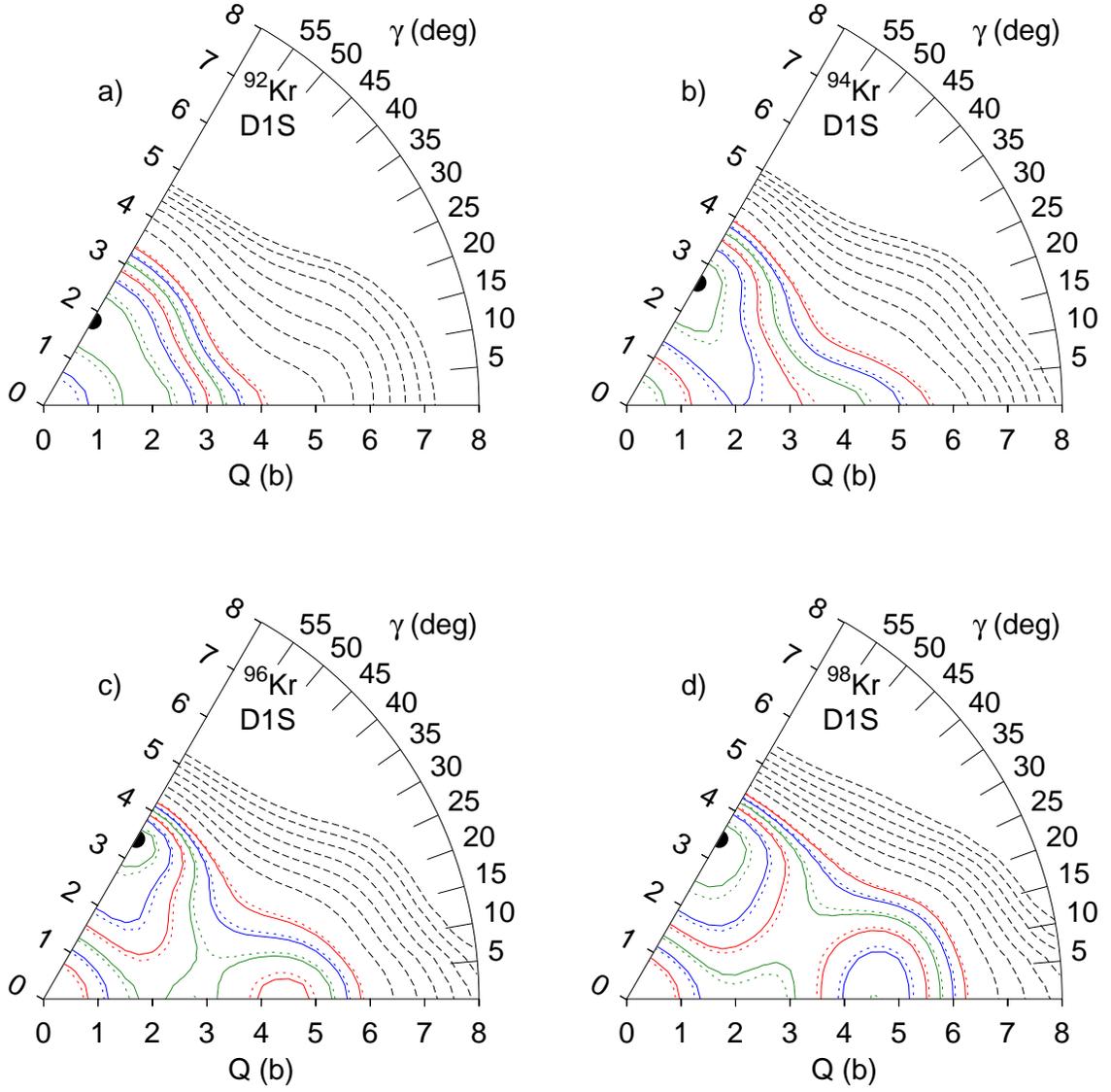}
\caption{(Color online) $Q-\gamma$ planes for $^{92}$Kr (a),
$^{94}$Kr (b), $^{96}$Kr (c), and $^{98}$Kr (d) with the Gogny-D1S
EDF. The absolute minimum is marked by a bullet. The full contour lines
correspond to energies $\epsilon_C$ (relatives to the absolute minimum)
of 0.25, 0.75, 1.25, 1.75, 2.25 and 2.75 MeV. Close to those full line
contours other contour lines corresponding to $\epsilon_C+0.1$ MeV are 
also depicted. The purpose of these dotted contour lines is to give
the direction of increasing energy as well as a visual idea of the 
corresponding slope. Finally, the dashed contour lines corresponding
to energies from 4 MeV up to 10 MeV in steps of 1 MeV are depicted
to mark the region where the potential energy starts to grow rapidly.}
\label{fig_kr_tri}
\end{figure*}

Finally, we also include results for Kr isotopes, measured in
Ref. \cite{keim}. The isotope shifts for Kr isotopes do not show any
abrupt change, but increase smoothly with the neutron number N. To
understand this behavior in Kr nuclei, at variance with the abrupt
change observed in the heavier neighboring isotopic chains, we have
studied the  evolution of the corresponding Potential Energy 
Surfaces (PESs) (i.e., $Q-\gamma$ energy contour plots) in even-even
Kr isotopes with the help of constrained HFB calculations along 
the lines discussed in Refs. \cite{ours4,ours5}.
We find that the lighter Kr isotopes present very shallow
PESs centered at the spherical shapes. Heavier Kr nuclei become
gradually oblate in their ground states with prolate configurations
at higher energies. Therefore, the oblate configurations are
stabilized in Kr isotopes and we do not find any transition to
prolate shapes. The result is a smooth variation of the ground-state
structure and, as a consequence, of the isotope shifts. To illustrate
this point we show in Fig. \ref{fig_kr_tri} the triaxial landscapes
for even-even  Kr isotopes in the vicinity of $N=60$. We can see that
the ground state in $^{92}$Kr (a) is oblate with a very soft variation in
$Q$ and especially in the $\gamma$ direction. The nucleus $^{94}$Kr (b)
develops an oblate ground state, which is again $\gamma$-soft.
A prolate saddle point is also apparent. The heavier isotopes  
$^{96}$Kr ($N=60$) (c) and  $^{98}$Kr ($N=62$) (d) develop two well defined
oblate and prolate minima, separated by energy barriers in the
$Q$ and  $\gamma$ degrees of freedom. The ground state corresponds
always to the oblate shapes, while the prolate shapes
appear about 1 MeV higher in energy.

In general, our theoretical calculations describe successfully this 
phenomenology. The smooth behavior observed in Kr isotopes is a
consequence of the stabilization of the oblate shapes along the
isotopic chain. The sudden increase of the charge radii at $N=60$
observed in Rb, Sr, Y, and Zr isotopes is explained by the 
oblate-to-prolate shape transition. Finally, the tendency to
a smooth behavior observed again in Nb and Mo isotopes is explained
by the onset of a region of triaxiality.

\section{Conclusions}
\label{CONCLU}

In this work we have studied the shape evolution in odd-$A$ Yttrium
and Niobium isotopes from microscopic selfconsistent Gogny-EDF HFB-EFA 
calculations. We have also analyzed the isotopic evolution of the
one-quasiproton configurations, comparing the predictions of the
parametrizations D1S and D1M of the Gogny-EDF and  demonstrating the
robustness of our calculations.

Two-neutron separation energies, charge radii, and the spin-parity of the
ground states have been analyzed in a search for signatures of shape
transitions. We have found that the charge radii and the spin-parity
of the ground states are very sensitive to shape changes in these
isotopes. In addition, the signatures found are all consistent to each
other. We have also shown that the quality of the spectroscopic
results obtained with the recent incarnation D1M of the Gogny-EDF
is comparable to the one obtained with the standard
parametrization D1S. We conclude that both D1M and D1S parametrizations
reproduce, at least qualitatively, the main features observed in the
isotopic trends of the considered neutron-rich isotopes. 

The observed spin-parity of the ground states exhibits a sudden
change between $N=58$ and $N=60$ isotopes in both isotopic chains.
This is a clear signature of a nuclear structure transition, which 
is correlated to the observation of a similar change in the charge
radii. According to our calculations this structural change is
explained by a shape transition taking place at this neutron number.
The relevant feature to stress is that the jump in both spin-parity 
and charge radius is a signature of a shape transition, where the
systematics is suddenly broken to a new situation. In the case of Y
isotopes, the experimental spin-parities are well reproduced, except
in the case of $N=58$, where the calculation predicts a transition going
across an oblate configuration, apparently not seen experimentally.
In the case of Nb isotopes, the predicted transition is to oblate 
states, while the experimental spin-parities of heavier isotopes do
not support these results. Since this is a region of an emergent
triaxiality, as it has been shown in neighboring even-even nuclei,
this is a plausible explanation of the discrepancy. Finally, one should
also take present that the experimental spin-parity assignments in the
heavier isotopes of both chains are either uncertain or taken from
systematics, and therefore, subject to revision.

We have performed a systematic study of the behavior of the charge
radii in the region of neutron-rich isotopes from Kr ($Z=36$) up to
Mo ($Z=42$) to stress the manifest enhanced sensitivity of these
observables to shape transitions. Comparison with the available data from
laser spectroscopy demonstrates the quality and robustness of the Gogny-HFB
description that is able to reproduce the main features of this behavior
and offers an intuitive interpretation in terms of sharp or soft shape
transitions, as well as triaxiality.

The experimental information available in neutron-rich Y and Nb isotopes,
and in general in this mass region, is still incomplete. It would be
desirable to extend the experimental programs for masses, charge radii, and
spectroscopic measurements to these exotic regions at existing facilities
like the tandem Penning trap mass spectrometer ISOLTRAP \cite{mukherjee} 
at the On-Line Isotope Mass Separator (ISOLDE) facility at CERN and the
Ion Guide Isotope Separator On-Line (IGISOL) facility \cite{jokinen}
at the University of Jyv\"askyl\"a, or at future ones like 
the precision measurements of very short-lived nuclei using an
advanced trapping system for highly-charged ions(MATS) and
laser spectroscopy (LaSpec) \cite{rodriguez} at the Facility for
Antiproton and Ion Research (FAIR), where we can learn much
about structural evolution in nuclear systems.

Although the present theoretical approach explains reasonably well the
basic features observed in the data,
theoretical efforts should be still pushed forward by improving the
formalism by including triaxial degrees of freedom in odd-$A$ isotopes, 
in particular in those regions where this can be an issue, or by dealing
with the odd systems with exact blocking treatments.
The quality of our mean field description could be also improved from
configuration mixing calculations in the spirit of the GCM
 with the quadrupole moment as collective coordinate.
It is already known \cite{bender08} that such a configuration mixing
reduces the jump in the predicted $S_{2n}$, as compared to pure mean-field
predictions, when crossing shell closures, thus improving the agreement
with experiment. This could be particularly relevant for the light
isotopes considered in the present study, where the spherical minima are
rather shallow. For heavier isotopes, the two minima, oblate and prolate,
are separated by spherical barriers of about 3 MeV and appear typically
about 1 MeV apart. The effect here is not expected to be significant
because a single shape would be enough to account for the properties
studied. 

\noindent {\bf Acknowledgments}

This work was supported by MICINN (Spain) under research grants 
FIS2008-01301, FPA2009-08958, and FIS2009-07277, as well as by 
Consolider-Ingenio 2010 Programs CPAN CSD2007-00042 and MULTIDARK 
CSD2009-00064. Author R.R acknowledges valuable suggestions 
from  Prof. J.\"Aysto, I. Moore, and the experimental teams of the
University of Jyv\"askyl\"a (Finland).

\begin{figure*}[h]
\centering
\includegraphics[width=150mm]{nb}
\caption{(Color online) Same as in Fig. \ref{fig_y}, but for Niobium isotopes.}
\label{fig_nb}
\end{figure*}

\end{document}